\documentclass[a4paper,11pt]{article}
\pdfoutput=1
\usepackage{jheppub}
\usepackage{textcase}
\usepackage{graphicx,epsfig,psfrag,amssymb,hyperref}
\usepackage{multirow}
\usepackage{color,graphicx,epsfig,psfrag,amsmath,empheq}
\usepackage[dvipsnames]{xcolor}
\usepackage{bm}
\usepackage{mathrsfs,amsfonts,soul,color}
\usepackage[caption=false]{subfig}
\usepackage{hepunits}
\usepackage{enumitem}
\usepackage{placeins}
\usepackage{textcomp}

\usepackage{float}
\usepackage[toc,page]{appendix}
\usepackage[ruled,vlined]{algorithm2e}
\usepackage{wrapfig}
\SetKwInOut{Input}{Input}
\SetKwInOut{Output}{Output}
\SetKwFunction{Fforward}{Forward}
\SetKwFunction{Fbackward}{Backward}
\SetKwProg{Fn}{function}{:}{}
\DontPrintSemicolon

\begin{document}


\title{Enhancing searches for resonances with machine learning and moment decomposition} 

\author{Ouail Kitouni,$^{a,e}$}
\author{Benjamin Nachman,$^{b,c}$}
\author{Constantin Weisser,$^{a,e}$}
\author{and Mike Williams$^{a,d,e,f}$}

\affiliation{
\begin{scriptsize}
\phantom{ }\hspace{-0.12in}$^a$Laboratory for Nuclear Science, Massachusetts Institute of Technology, \!Cambridge, \!MA 02139, \!USA \\
\phantom{ }\hspace{-0.12in}$^b$Physics Division, Lawrence Berkeley National Laboratory, Berkeley, CA 94720, USA \\
\phantom{ }\hspace{-0.12in}$^c$Berkeley Institute for Data Science, University of California, Berkeley, CA 94720, USA \\
\phantom{ }\hspace{-0.12in}$^d$Statistics and Data Science Center, Massachusetts Institute of Technology, \!Cambridge, \!MA 02139, \!USA \\
\phantom{ }\hspace{-0.12in}$^e$The NSF AI Institute for Artificial Intelligence and Fundamental Interactions \\
\phantom{ }\hspace{-0.12in}$^f$School of Physics and Astronomy, Monash University, Melbourne, Victoria 3168, Australia
\end{scriptsize}
}

\emailAdd{kitouni@mit.edu}
\emailAdd{bpnachman@lbl.gov}
\emailAdd{weisser@mit.edu}
\emailAdd{mwill@mit.edu}

\abstract{
A key challenge in searches for resonant new physics is that classifiers trained to enhance potential signals must not induce localized structures.  Such structures could result in a false signal when the background is estimated from data using sideband methods.  A variety of techniques have been developed to construct classifiers which are independent from the resonant feature (often a mass).  Such strategies are sufficient to avoid localized structures, but are not necessary.  We develop a new set of tools using a novel moment loss function (Moment Decomposition or \textsc{MoDe}) which relax the assumption of independence without creating structures in the background.  By allowing classifiers to be more flexible, we enhance the sensitivity to new physics without compromising the fidelity of the background estimation.
}

\maketitle

\section{Introduction}
\label{sec:intro}

Searching for new phenomena associated with localized excesses in otherwise featureless spectra, often referred to as bump hunting, is one of the most widely used approaches in particle and nuclear physics, dating back at least to the discovery of the $\rho$ meson~\cite{Button:1962bnf}, and used continuously since, including recently in the discovery of the Higgs boson~\cite{Aad:2012tfa,Chatrchyan:2012ufa}.  In the present day, such searches reach the multi-TeV scale~\cite{Sirunyan:2019vgj,Aad:2019hjw} and span high energy particle and nuclear physics experiments~\cite{Aaij:2020ikh,Adam:2020zsu,Acharya:2020jil,Adrian:2018scb,McCracken:2015coa,Ablikim:2019rpl,BelleII:2020fag}.  A key feature of these searches is that they are relatively background model agnostic since sidebands in data can be used to estimate the background under a potential localized excess.  These sideband fits are possible because the background data can be well-approximated either with simple parametric functions or smooth non-parametric techniques such as Gaussian processes~\cite{Frate:2017mai}.

Sideband methods for background estimation are often combined with relatively simple and robust event selections in order to ensure broad coverage of new physics model space.  However, there is a growing use of modern machine learning techniques to enhance signal sensitivity~\cite{Larkoski:2017jix,Guest:2018yhq,Albertsson:2018maf,Radovic:2018dip,Bourilkov:2019yoi}.  For example, both ATLAS~\cite{Aaboud:2018psm} and CMS~\cite{Sirunyan:2020lcu} have developed machine-learning-based $W$ jet taggers that improve the sensitivity of searches involving Lorentz-boosted and hadronically decaying $W$ bosons.   Boosted electroweak bosons are common in searches for models with a significant mass hierarchy between the primary resonance mass and the $W$ boson mass~\cite{Aad:2019fbh,Aad:2020ddw,Sirunyan:2019jbg,Sirunyan:2019vgt,Aad:2020tps,Sirunyan:2018fuh,Aad:2020ldt,Sirunyan:2019quj,Sirunyan:2017isc,Aad:2020hzm,Aaboud:2017ecz,Aad:2020cws}, and boosted $W$-like particles are a feature of searches for low-mass dark matter mediators~\cite{Aaboud:2018zba,Sirunyan:2017dnz,Sirunyan:2018ikr,Sirunyan:2019sgo,Sirunyan:2019vxa,ATLAS-CONF-2018-052,Sirunyan:2017dgc}.    

A key challenge with complex event selections like those involved in boosted $W$ tagging is that they can invalidate the smoothness assumption of the background.  In particular, if classifiers can infer the mass of the parent resonance, then selecting signal-like events will simply pick out background events with a reconstructed mass near the target resonance mass.  Many techniques have been developed that modify or simultaneously optimize classifiers so that their responses are independent of a given resonance feature~\cite{NIPS2017_6699,Dolen:2016kst,Moult:2017okx,Stevens:2013dya,Shimmin:2017mfk,Bradshaw:2019ipy,ATL-PHYS-PUB-2018-014,DiscoFever,Xia:2018kgd,Englert:2018cfo,Wunsch:2019qbo,Rogozhnikov:2014zea,10.1088/2632-2153/ab9023,clavijo2020adversarial,Kasieczka:2020pil,Chang:2017kvc,Clavijo:2020mua,Sirunyan:2019nfw,Stevens:2013dya,Rogozhnikov:2014zea}.  For machine learning classifiers, the proposed solutions include modifications to loss functions that implicitly or explicitly enforce independence. 
These methods have been successfully deployed in bump hunts; see, {\em e.g.}, Refs.~\cite{Sirunyan:2017dnz,Sirunyan:2017dgc,Sirunyan:2017nvi,Aaboud:2018zba,Sirunyan:2018ikr,Sirunyan:2018gdw,Sirunyan:2018sgc,Sirunyan:2019sgo,Sirunyan:2019jbg,Sirunyan:2019vxa,Aad:2020cws,Sirunyan:2020hwz,Aaij:2020ypa,Aaij:2020hon,Aaij:2019mhf,Aaij:2018xpt,Aaij:2015tna,Aaij:2017eru,Aaij:2016vsy}.
A variety of similar proposals under the monikers of domain adaptation and fairness have been proposed in the machine learning literature (see e.g. Ref.~\cite{DBLP:journals/corr/EdwardsS15,JMLR:v17:15-239} and Ref.~\cite{,mehrabi2019survey,alex2018frontiers}).

Ensuring that a classifier is independent from a given resonant feature is sufficient for mitigating sculpting, but it is not necessary.  The original requirement is simply that a selection using the classifier does not introduce localized features in the background spectrum, which is a much looser requirement than enforcing independence.  For example, if a classifier has a linear dependence on the resonant feature, then there would be a strong correlation.  However, a threshold requirement on such a classifier would not sculpt any bumps in the background-only case.  This example motivates a new class of techniques that allow classifiers to depend on the resonant feature in a controlled way.  In the limit that constant dependence is required, then the classifier and the resonant feature will be independent.  The advantage of relaxing the independence requirement is that the resulting classifiers can achieve superior performance because they are allowed to be more flexible.  

In this article, we present a new set of tools that allow for controlled dependence on a resonant feature. 
This new approach is called \textit{Moment Decomposition} (\textsc{MoDe}).
Using \textsc{MoDe}, analysts can require independence, linear dependence, and quadratic dependence. 
In addition, analysts can place bounds on the slope of the linear dependence, and restrict quadratic dependence to be monotonic.
Extending \textsc{MoDe} to allow for arbitrarily higher-order dependence is straightforward. 
This article is organized as follows.  Section~\ref{sec:methods} briefly reviews existing decorrelation methods and then introduces \textsc{MoDe}.  Numerical results using a simplified model and a physically motivated example are presented in Sec.~\ref{sec:results}. Finally, we present conclusions and outlook in Sec.~\ref{sec:conclusion}.

\section{Methods}
\label{sec:methods}

\subsection[Existing Decorrelation Methods]{Existing decorrelation methods}

We will consider the binary classification setting in which examples are given by the triplet $(X,Y,M),$ where $X\in \mathcal{X}$ is a feature vector, $Y\in\mathcal{Y}:=\{0,1\}$ is the target label, and finally, $M\in \mathcal{M}$ is the resonant feature (or protected attribute) whose spectrum will be used in the bump hunting. 
Throughout this article, we take $M$ to be mass, though it could be any feature. 
The feature vector $X$ can either contain $M$ directly as one of its elements or contain other features that are arbitrarily indicative of $M$. 
We are interested in finding a mapping $ f : \mathcal{X} \rightarrow \mathcal{S}$ where $ s\in \mathcal{S}$ are scores used to obtain predictions $\hat{y}\in \mathcal{Y}$ with the additional constraint that $f$ be conditionally independent of (or uniform with) $M$ in the sense that
\begin{align}
    \label{eqn:uniform}
    p(f(X)=s|M=m,Y=y) = p(f(X)=s|Y=y)  \; \forall \; m \in \mathcal{M} \; \text{and} \; \forall \; s \in \mathcal{S}, 
\end{align}
for one or more values $y$, 
although typically, 
Eq.~\eqref{eqn:uniform} is required only for the background.

Existing decorrelation methods used in particle physics that simultaneously train a classifier $f(x):\mathbb{R}^n\rightarrow [0,1]$ and decorrelate from a resonant feature $m$ use the following loss function:
\begin{align}
\label{eq:loss}
    \mathcal{L}[f(x)]=\sum_{i\in\text{S}}L_\text{class}(f(x_i),1)+\sum_{i\in\text{B}}w(m_i) \,L_\text{class}(f(x_i),0)+\lambda \sum_{i\in\text{B}}L_\text{decor}(f(x_i),m_i)\,,
\end{align}
where $S=\{i | \, y_i=1\}$ and $B=\{i | \, y_i=0\}$ denote signal and background, respectively, $L_\text{class}$ is the usual classification loss such as the binary cross entropy $L_\text{BCE}(f(x),y)=y\log(f(x))+(1-y)\log(1-f(x))$, $w$ is a weighting function, $\lambda$ is a hyperparameter, and $L_\text{decor}$ generically denotes some form of decorrelation loss.  Standard classification corresponds to $w(m)=1$ and $\lambda=0$.  Decorrelation methods include:

\begin{itemize}[topsep=.6em,leftmargin=1.0em,itemindent=0.2em,itemsep=0.4em]
    \item Planing~\cite{Chang:2017kvc,1511.05190}: $\lambda=0$ and $w(m_i)\approx p_S(m)/p_B(m)$ so that the marginal distribution of $m$ is non-discriminatory after the reweighting.
    \item Adversaries~\cite{NIPS2017_6699,Shimmin:2017mfk,Englert:2018cfo,Clavijo:2020mua}:  $w(m)=1$, $\lambda<0$, and $L_\text{decor}$ is the loss of a second neural network (adversary) that takes $f(x)$ as input and tries to learn some properties of $m$ or its probability density.
    \item Distance Correlation (DisCo)~\cite{DiscoFever,Kasieczka:2020pil}: $w(m)=1, \lambda>0$, and the last term in Eq.~\eqref{eq:loss} is the \textit{distance correlation}~\cite{szekely2007, szekely2009, SzeKely:2013:DCT:2486206.2486394,szekely2014} between $f(x)$ and $m$ for the background.\footnote{Technically, the term $L_\text{decor}$ is applied at the level of a batch because it requires computing expectation values over pairs of events.}
    \item Flatness~\cite{Rogozhnikov:2014zea}: $w(m)=1$, $\lambda>0$, and $L_\text{decor}=\sum_m b_m \int |F_m(s)-F(s)|^2\, {\rm d}s$ where the sum runs over mass bins, $b_m$ is the fraction of candidates in bin $m$, $F$ is the cumulative distribution function, and $s=f(x)$ is the classifier output.
\end{itemize}
Decorrelation methods have proven to be useful additions to the toolkit of the bump hunter. 

\subsection[Moment Decorrelation]{Moment decorrelation}

First, we will derive a new decorrelation method based on moments. While this technique achieves state-of-the-art decorrelation performance, along with being robust, simple, and fast, its true value is that it is trivially extended to allow for controlled dependence beyond just decorrelation. 

We begin by noting that the uniformity constraint in Eq.~\eqref{eqn:uniform} can be written in terms of the conditional cumulative distribution function (CDF) of scores at $s$, $F(s|M,Y)$, as 
\begin{align}
\label{eq:uniform_cdf}
    F(f(X)=s|Y=y) = F(f(X)=s|M=m,Y=y) \; \forall \; m \in \mathcal{M} \; \text{and} \; \forall \; s \in \mathcal{S}. 
\end{align}
This is the same observation that lies at the heart of the flatness loss defined in  Ref.~\cite{Rogozhnikov:2014zea}.
Here, we will consider the conditional CDFs in bins of mass and only on the background, which allows us to adopt the following more compact notation
\begin{align}
    F(f(X)=s|M=m,Y=y) \to F_m(s),
\end{align}
where now $m$ is discrete and indexes the mass bins. 
We leave the exploration of similar unbinned approaches for future work. 
Furthemore, we assume that some transformation is performed on $m$ such that $\mathcal{M} \to [-1,1]$. This could be a simple linear transformation but does not have to be, discussion on this point is provided later. 

The uniformity constraint of Eq.~\eqref{eq:uniform_cdf} can be imposed on the learned function by defining the decorrelation loss using\footnote{We do not presume to know what the analyst is going to do with the trained model; therefore, we weight all score values equally seeking to achieve decorrelation for any score threshold. If additional information is available about how the model will be used, another choice of weighting function of the form d$s\to w(s){\rm d}s$ could be used instead, though it would be important to ensure that the functional derivative of the {\sc MoDe} loss can still be calculated precisely; see Sec.~\ref{sec:comp}.}
\begin{align}
\label{eq:lossm0}
    L_{\rm decor} \to L_{\rm\sc MoDe}^0 \equiv \sum_m  \int | F_m(s) - F_m^0(s)|^2 {\rm d}s.
\end{align}
Here, $F^0_m$ is based on the 0$^{\rm th}$ Legendre\footnote{Any choice of orthogonal polynomials would work here.} moment of $F_m(s)$ in $m$, $c_0$, and polynomial, $P_0(x)=1$, as
\begin{align}
  \label{eq:F0}
F^0_m(s)  = c_0(s) P_0(\tilde{m}) = \frac{1}{2} \int_{-1}^{+1} P_0(m') F(s|m') {\rm d}m' \approx \frac{1}{2} \sum_{m'} \Delta_{m'} F_{m'}(s),
\end{align}
where $\Delta_m$ denotes the width of bin $m$, and $\tilde{m}$ is its  central mass value. 
Note that the loss in Eq.~\eqref{eq:lossm0} is clearly minimized when
\begin{align}
    F_m(s) = F_m^0(s) = c_0(s) \; \forall \; m,
\end{align}
which implies that Eq.~\eqref{eq:uniform_cdf} holds and $f(X)$ and $M$ are indeed independent. 
Note that in the limit that all bins have equal width and occupancy, it is straightforward to show that the loss function in Eq.~\eqref{eq:lossm0} is the same as the flatness loss of Ref.~\cite{Rogozhnikov:2014zea}; however, when the underlying background distribution is highly nonuniform, these loss functions are drastically different resulting in {\sc MoDe} outperforming Ref.~\cite{Rogozhnikov:2014zea} in such cases. 

\subsection[Beyond Decorrelation: Moment Decomposition]{Beyond decorrelation: Moment decomposition}

We will now generalize moment decorrelation to allow for controllable mass dependence in the form of an $\ell^{\rm th}$ order polynomial, where $\ell$ is a hyperparameter chosen by the analyst. 
The generalized {\sc MoDe} loss is given by
\begin{align}
\mathcal{L}[f] = L_{\mathrm{class}} + \lambda L^{\ell}_{\mathrm{\textsc{MoDe}},}    
\end{align}
where 
\begin{align}
\label{eq:loss-mode}
    L_{\rm\sc MoDe}^{\ell} \equiv \sum_m  \int | F_m(s) - F_m^{\ell}(s)|^2 {\rm d}s. 
\end{align}
Here, $F^0_m$ in Eq.~\eqref{eq:lossm0} has been replaced by 
\begin{align}
\label{eq:Fmell}
    F_m^{\ell}(s) = \sum_{l=0}^{\ell} c_l(s) P_l(\tilde{m}),
\end{align} 
and the Legendre moments are given by
\begin{align}    
\label{eq:ci}
   c_l(s) = \left[ \frac{2 l + 1}{2} \right] \int_{-1}^1 P_l(m') F(s|m') {\rm d}m' \approx \left[ \frac{2 l + 1}{2} \right] \sum_{m'} \Delta_{m'} P_l(\tilde{m}') F_{m'}(s).
\end{align}
We note that setting $\ell = 0$ reduces the generalized {\sc MoDe} loss of Eq.~\eqref{eq:loss-mode} down to the moment decorrelation of Eq.~\eqref{eq:lossm0}. 

The {\sc MoDe} loss in Eq.~\eqref{eq:loss-mode} is optimal when $F_m(s) = F^{\ell}_m(s) \; \forall \; m, s$, which clearly occurs when the mass dependence of the classifier is at most an $\ell^{\rm th}$ order polynomial. 
For example, taking $\ell = 0$ drives the classifier to be independent of mass.
More interestingly, choosing $\ell = 1$ allows for a linear mass dependence, $\ell = 2$ quadratic dependence, {\em etc.} 
Furthermore, making the replacement\footnote{The hyperbolic tangent function has several beneficial properties which motivate its usage here---its range is $(-1,1)$, it is differentiable, monotonic, and odd---although other functions could be substituted.} 
\begin{align}
    c_1(s) \to c_1^{\rm max}c_0(s) \tanh{\left(\frac{c_1(s)}{c_1^{\rm max}c_0(s)} \right)}
\end{align}
in Eq.~\eqref{eq:Fmell} places an upper limit $c_1^{\rm max}c_0(s)>0$ on the magnitude of the linear slope (the first Legendre moment is the coefficient of the $\tilde{m}$ term), allowing the analyst to control this aspect of the mass dependence through a hyperparameter, $c_1^{\rm max}$.
In addition, for the case where $\ell = 2$ is selected, it is straightforward to show that as long as ${3 |c_2(s)| \leq |c_1(s)|}$ the derivative of $F^{\ell}_m(s)$ is nonzero on $(-1,1)$. 
Therefore, making the replacement 
\begin{align}
    c_2(s) \to \frac{c_1(s)}{3} \tanh{\left(\frac{3c_2(s)}{c_1(s)} \right)}
\end{align}
in Eq.~\eqref{eq:Fmell} results in monotonic mass dependence. 
This option can be turned on or off in {\sc MoDe}, and can be used in conjunction with $c_1^{\rm max}$ if desired. Finally, controlled higher-order mass dependence can be achieved  by extending these ideas to larger $\ell$ values. 

\subsection[Computational Details]{Computational details}
\label{sec:comp}

Computing the {\sc MoDe} loss and its gradient is straightforward using a few approximations.
At the batch level, 
\begin{align}
\label{eq:Fmapprox}
F_m(s) \approx \frac{1}{n_m} \sum_{i=1}^{n} \Theta(s - s_i) \delta_{m,m_i},    
\end{align}
where $n$ is the number of samples in the batch, $n_m$ is the number of samples in bin $m$, $s_i \equiv f(x_i)$ is the score of sample $i$, and $\Theta$ is the Heaviside function: $\Theta(x)=1$ if $x>0$ and $\Theta(x)=0$ otherwise. 
Minimizing the loss function requires calculating the functional derivative of $L_{\sc\rm MoDe}^{\ell}$ with respect to $f$. 
This requires specifying how the {\sc MoDe} loss changes due to variations of the score of each sample in the batch: 
\begin{align}
\delta L_{\sc\rm MoDe}^{\ell} =  \delta s_i \sum_m \int 2 \left[ F_m(s) - F^{\ell}_m(s) \right] \left[ \frac{\partial F_{m}}{\partial s_i} - \frac{\partial F^{\ell}_{m}}{\partial s_i}  \right] {\rm d}s, 
\end{align}
where from Eq.~\eqref{eq:Fmapprox} 
\begin{align}
\label{eq:dFmds}
\frac{\partial F_{m}}{\partial s_i} \approx -\frac{1}{n_{m_i}} \delta(s - s_i) \delta_{m,m_i}.
\end{align}
In addition, using this result, along with Eqs.~\eqref{eq:Fmell} and \eqref{eq:ci}, we obtain
\begin{align}
    \frac{\partial F^0_{m}}{\partial s_i} & \approx \frac{1}{2}\Delta_{m_i} \frac{\partial F_{m_i}}{\partial s_i} = -\frac{\Delta_{m_i}}{2n_{m_i}} \delta(s - s_i), \\
    \frac{\partial F^1_{m}}{\partial s_i} & \approx \frac{\partial F^0_{m}}{\partial s_i} + \frac{3}{2} \tilde{m} \cdot \tilde{m}_i \Delta_{m_i} \frac{\partial F_{m_i}}{\partial s_i} = -\frac{\Delta_{m_i}}{2n_{m_i}} \delta(s - s_i)  \left[1 + 3 \tilde{m} \cdot \tilde{m}_i\right], \\
    \vdots \nonumber 
\end{align}
where the sum over mass bins in Eq.~\eqref{eq:ci} is no longer needed, since changes to the score for sample $i$ only affect the CDF in bin $m_i$. 
The factors of $\delta(s - s_i)$ eliminate the integral over $s$ resulting in relatively simple gradient terms, {\em e.g.}, for $\ell = 1$ we obtain
\begin{align}
\label{eq:L1partial}
    \delta L_{\sc\rm MoDe}^{1} \approx - \delta s_i \sum_m \frac{1}{n_{m_i}} \left[ F_m(s) - F^{1}_m(s) \right] \left[2 \delta_{m,m_i} - \Delta_{m_i}(1 + 3 \tilde{m} \cdot \tilde{m}_i) \right].
\end{align}
The fact that terms like Eq.~\eqref{eq:L1partial} do not depend on how the integral over $s$ is approximated yields high-precision gradients, which  is a big advantage when performing gradient descent.

The results in this subsection are easily generalized for weighted samples. 
The CDFs in Eq.~\eqref{eq:Fmapprox} become 
\begin{align}
    F_m(s) \approx \frac{1}{w_m} \sum_i^{n} w_i \Theta(s - s_i) \delta_{m,m_i},
\end{align}
where $w_i$ are the per-sample weights and
\begin{align}
    w_m = \sum_i^{n} w_i \delta_{m,m_i}
\end{align}
is the sum of the weights in bin $m$.
Equation~\eqref{eq:dFmds} then becomes
\begin{align}
    \frac{\partial F_{m}}{\partial s_i} \approx -\frac{w_i}{w_{m_i}} \delta(s - s_i) \delta_{m,m_i},
\end{align}
and updating the rest of the results follows accordingly.

Finally, we address the topic of scalability. 
While the optimization of the \textsc{MoDe} loss works well stochastically with few examples every step, its performance increases greatly with larger batch sizes.	
This is not surprising due to the {\em global} nature of the \textsc{MoDe} constraint.
Fortunately, all of the calculations scale well with batch size (see Appendix~\ref{app:optim} for time and memory performance as functions of the number of inputs).
Most computational costs occur in the forward direction, where \textsc{MoDe} scales linearly with the number of inputs $n$ (batch size) and the number of steps chosen for the integral in $s$, $n_s$. 
In addition, dynamic binning sorts $m_i$ and reindexes $s_i$ are required, and so $\textsc{MoDe}$ runs in $\mathcal{O}(n_s\times n + n\log n)$ time. 
In the forward direction, we also compute and cache the residual $F_m(s_i)-\Tilde{F}_m(s_i)$ which is used in the backward pass.
Since the CDFs are evaluated at every $s_i$, this contributes an $\mathcal{O}(n\times n_m)$ component.  
In theory, this could be improved, {\em e.g.}, if the CDFs at $s_i$ were instead approximated using nearest neighbor interpolation.
Finally, \textsc{MoDe} takes $\mathcal{O}(n\times n_m)$  extra memory (beyond what is required to store the data) when calculating the gradients, since the CDF is computed for each input for each bin.
It is worth noting that, at particularly small batch sizes, \textsc{MoDe} might be susceptible to slow convergence due to mini-batch statistics not accurately reflecting the full-batch statistics. That is why we recommend using \textsc{MoDe} with a sizeable fraction of the full sample.


\section{Example Results}
\label{sec:results}
In this section, we will demonstrate how {\sc MoDe} performs on a simple model problem, and on the $W$-jet tagging problem used in the decorrelation studies of Ref.~\cite{DiscoFever,ATL-PHYS-PUB-2018-014}. 
All of the numerical results reported in this section are obtained using the PyTorch framework~\cite{NEURIPS2019_9015}. 

\subsection[Simple Model]{Simple Model}
\label{sec:toy}

\begin{figure}[b]
    \centering
    \includegraphics[width=1\textwidth]{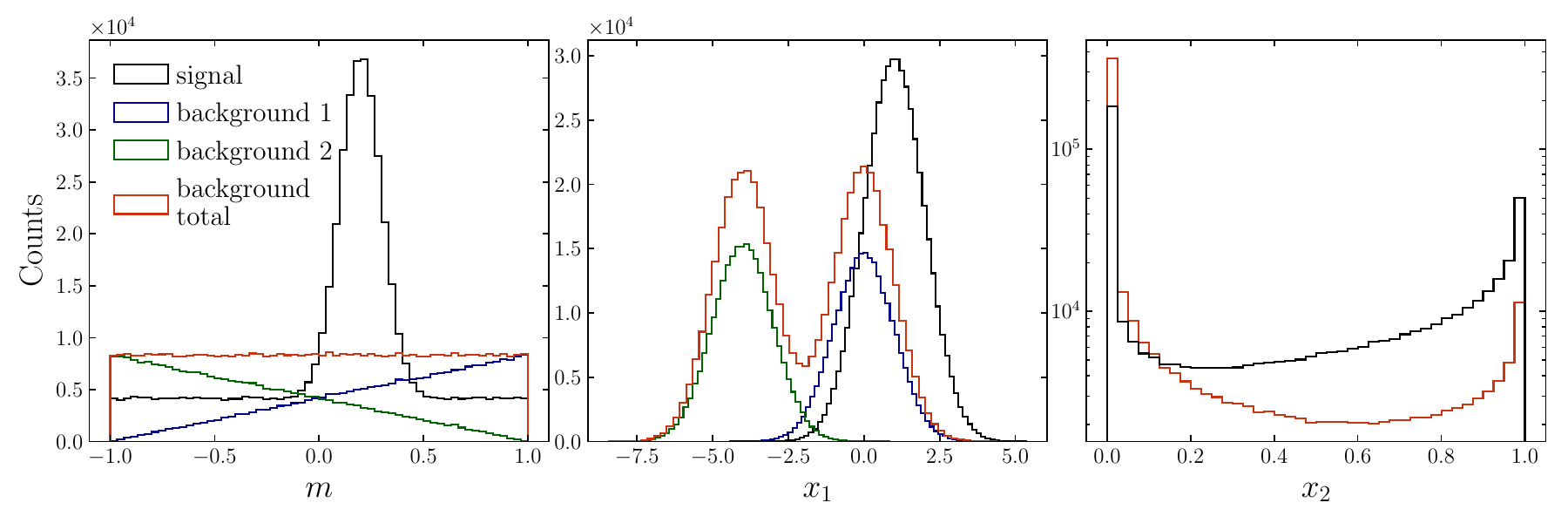}
    \vspace*{-10mm}
    \caption{Simple model distributions.}
    \label{fig:toy}
\end{figure}

We first consider a binary classification example composed of a signal and two types of background. 
Each sample $X \in \mathcal{X}$ has 2 features:
\begin{align}
    x_1 & \sim 
    \begin{cases}
    \mathcal{N} \big( 1,1\big) \quad & \quad \text{when} \; Y=1,\\
    \mathcal{N} \big( 0,1\big) \quad & \quad \text{when}\; Y=0 \;  \text{for background type 1},\\
    \mathcal{N} \big( -4,1\big) \quad & \quad \text{when} \; Y=0 \; \text{for background type 2},\\
    \end{cases}\\
    x_2 &= \exp\left[-\frac{(m-0.2)^2}{2\cdot 0.1^2}\right] \; \; \text{when} \; Y=0 \; \text{or } Y=1,
\end{align}
where $\mathcal{N}$ denotes the normal distribution. 
The mass, $m$, is drawn from $\mathcal{N}(0.2,0.1)$ and $\mathcal{U}(-1,1)$ at equal rates when $Y=1$. 
For the backgrounds, we sample a uniform random variable, $U \sim \mathcal{U}(0,1)$, then define $m = 1-2\sqrt{U}$ and $m = -1+2\sqrt{U}$ for background types 1 and 2, respectively.
This simple-model data is shown in Fig.~\ref{fig:toy}.

\begin{figure}[t]
    \centering
    \includegraphics[width=\textwidth]{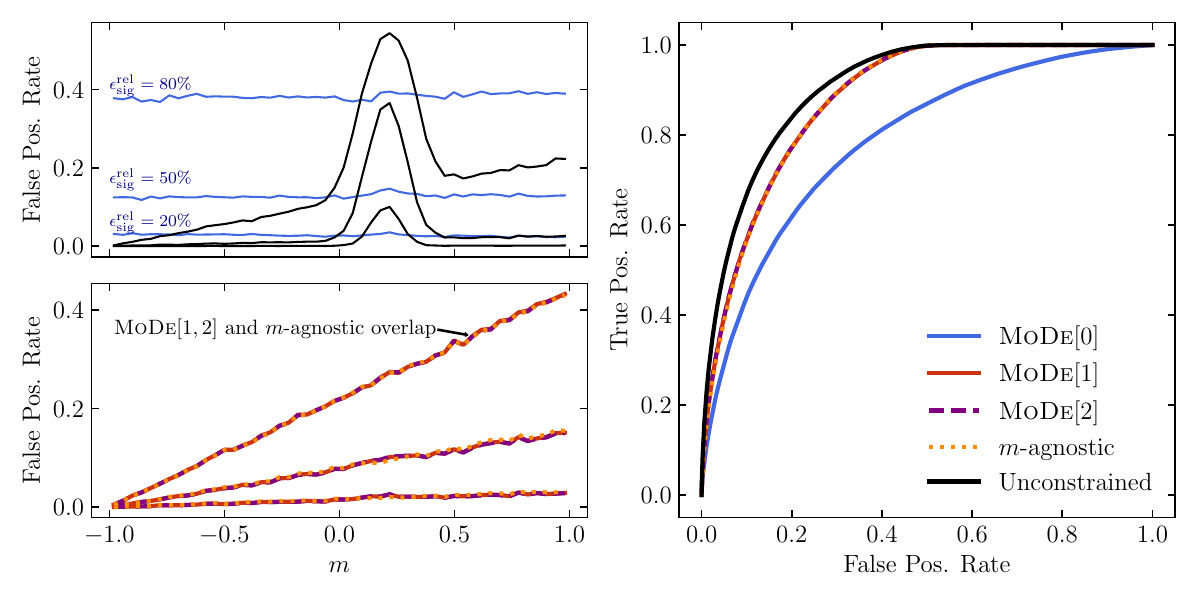}
    \vspace*{-10mm}
    \caption{\textbf{Left:} The false positive rate versus mass for various models
    at signal efficiencies $\epsilon^\text{rel}_\text{sig} = 80, 50, 20\%$ (each set of 3 identically colored and stylized lines correspond to the same model but with selection thresholds chosen to achieve the 3 desired signal efficiencies). The bottom panel shows that \textsc{MoDe}[1] and {\sc MoDe}[2] completely overlap with the $m$-agnostic model for this simple example, which is expected because the optimal classifier here has linear dependence on mass (see text). \textbf{Right:} ROC curves for {\sc MoDe}[0], {\sc MoDe}[1], and {\sc MoDe}[2] compared to the $m$-agnostic model and a model with unconstrained mass dependence. As in the left panel, we see that {\sc MoDe}[1], {\sc MoDe}[2], and the $m$-agnostic ROC curves are nearly identical because the optimal classifier has linear mass dependence in this simple example.}
    \label{fig:orders}
\end{figure}

In this scenario, an unconstrained classifier with sufficient capacity will learn the underlying mass distribution (due to the explicit mass dependence of $x_2$) and use it to discriminate between signal and background. 
Figure~\ref{fig:orders} shows how such a classifier favors regions near $m=0.2$, leading to extreme peak-sculpting in the background.  
It would be difficult to employ this classifier in a real-world analysis and obtain an unbiased signal estimator.  
Figure~\ref{fig:orders} also shows that {\sc MoDe}[0] successfully decorrelates the classifier response from mass producing a viable classifier for such an analysis.

In this simple example, we can easily choose to only use information not explicitly indicative of mass by removing $x_2$ from $\mathcal{X}$. 
Figure~\ref{fig:orders} shows that the resulting {\em mass agnostic} classifier is linearly correlated with mass.
Indeed, we ensured this via our choice of $x_2$, {\em i.e.}\ we configured this toy example such that the optimal classifier, the likelihood ratio, is linearly correlated with mass and obtained without the use of $x_2$. 
However, a classifier that enforces decorrelation must accept backgrounds 1 and 2 at equal rates to keep $p(s|m)$ flat, which as shown in Fig.~\ref{fig:orders} produces performance that is far from optimal. 
By relaxing the flatness constraint, {\sc MoDe}[1] is able to reject background 2 at a higher rate, while producing the expected linear dependence on mass.
This linear mass dependence, which will not sculpt out any fake peaks from the background, allows {\sc MoDe}[1] to achieve better classification power than is possible using decorrelation. 
In this case, it is able to achieve the same performance as the optimal mass-agnostic classifier, since the optimal performance here is linear.
Figure~\ref{fig:orders} shows that even though {\sc MoDe}[2] is given the freedom to find quadratic mass dependence, it also produces the same optimal linear mass dependence in this case.

\begin{figure}[t]
    \centering
    \includegraphics[width=.5\textwidth]{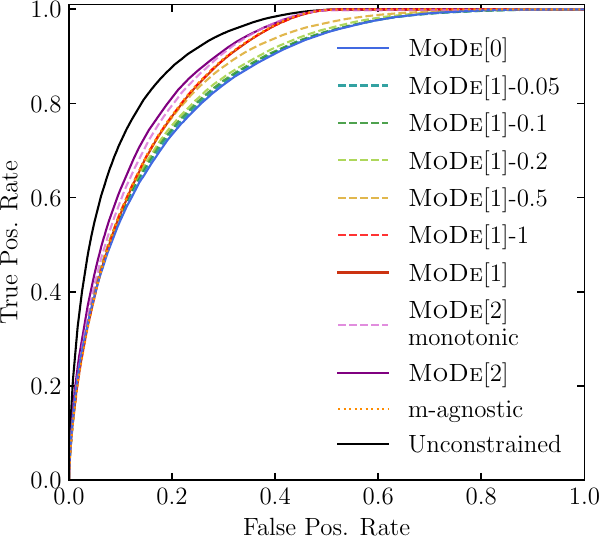}
    \caption{Same as the right panel of  Fig.~\ref{fig:orders} but with the simple-model modification of Eq.~\eqref{eq:toy2}. In addition, a monotonic version of {\sc MoDe}[2] and several versions of {\sc MoDe}[1] with constrained maximum slope values are also shown.}
    \label{fig:toy2}
\end{figure}

\begin{figure}[b]
    \centering
    \includegraphics[width=0.9\textwidth]{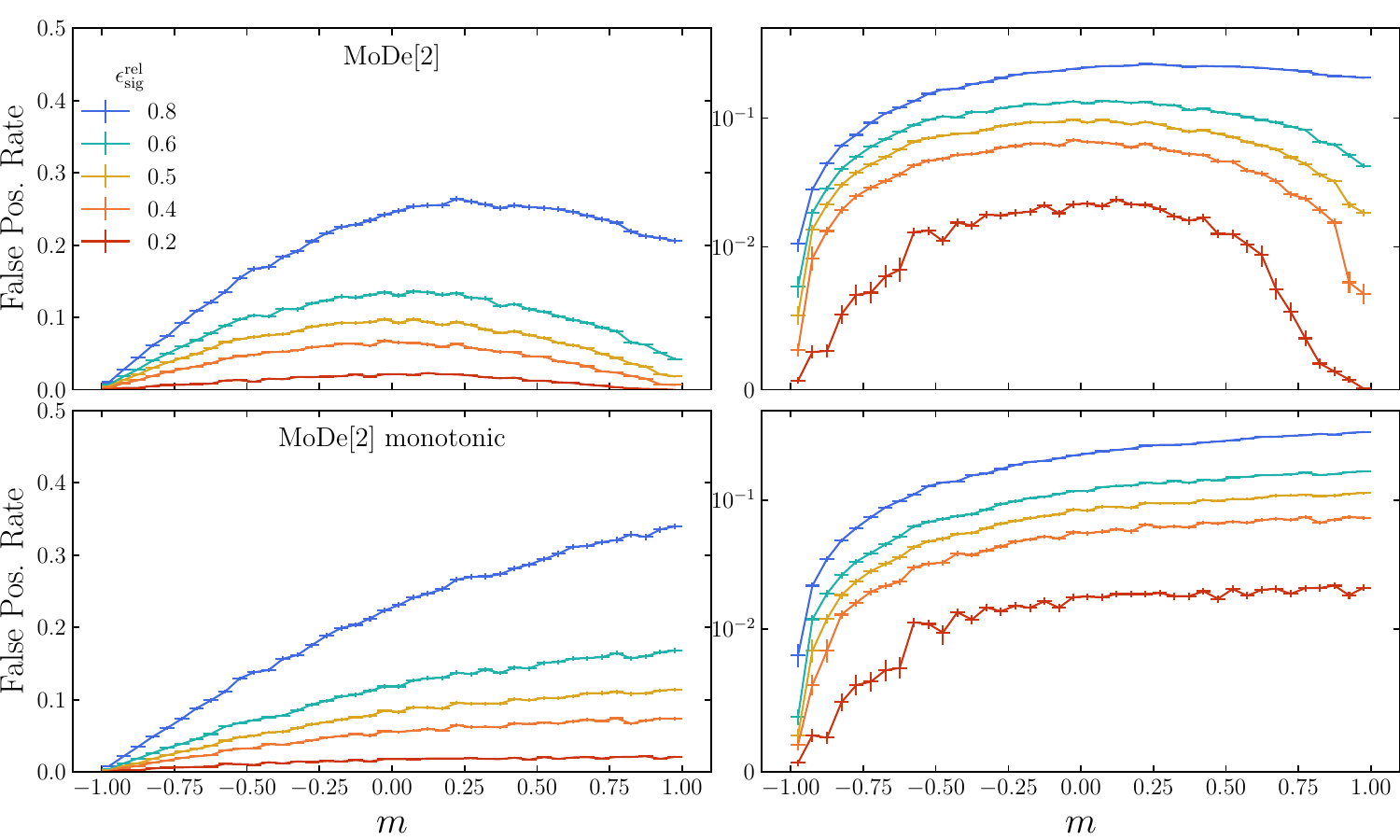}
    \caption{\textbf{Top:} False positive rate versus mass at various signal efficiencies for non-monotonic \textsc{MoDe}[2] on the modified simple-model example; see Eq.~\eqref{eq:toy2}. \textbf{Bottom:} False positive rate for monotonic \textsc{MoDe}[2]. {\em N.b.}, the right panels show the same curves as the left but on log scales.}
    \label{fig:monotonic}
\end{figure}

\begin{figure}[t]
    \centering
    \includegraphics[width=.95\textwidth]{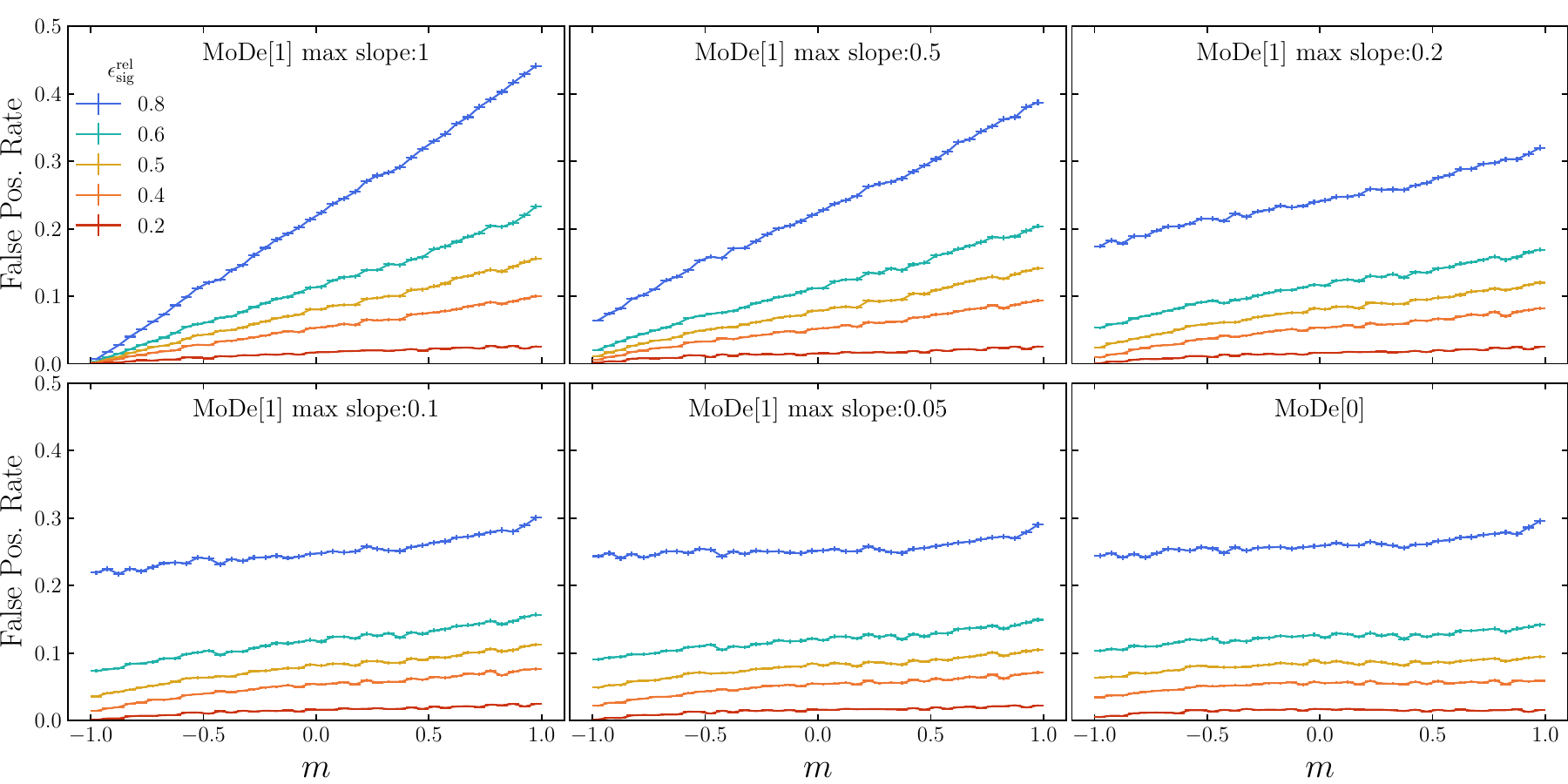}
    \vspace*{-5mm}
    \caption{Results for \textsc{MoDe}[1] on the modified simple-model example requiring various maximum slope values.}
    \label{fig:slopes}
\end{figure}

As discussed in Sec.~\ref{sec:methods}, the {\sc MoDe} package provides an even higher level of control over the response of a model, including allowing the analyst to define the maximum linear slope and to require that the quadratic dependence is monotonic. 
To demonstrate these features, we make the following minor change to the simple model:
\begin{align}
\label{eq:toy2}
    x_2 \to {\rm exp}(m) + 2m.
\end{align}
In this case, the optimal classifier is no longer linear.
Figure~\ref{fig:toy2} shows that here the additional freedom given to {\sc MoDe}[2] does improve the classification performance.
Figure~\ref{fig:monotonic} shows that the {\sc MoDe}[2] solution does indeed have quadratic mass dependence.
The {\sc MoDe}[2] response is not monotonic by default, but we also show in Fig.~\ref{fig:monotonic} that we can apply such a constraint.
As can be seen in Fig.~\ref{fig:toy2}, there is a small decrease in classification performance; whether this is acceptable is a problem-specific decision left to the analyst.
Finally, Fig.~\ref{fig:slopes} shows that the analyst can exert full control over the maximum linear slope. 
This could be desirable in cases where the signal mass is not known, and similar---but not necessarily equivalent---performance across the mass range is viewed as beneficial. 

\clearpage

\subsection[Boosted Hadronic W Tagging]{Boosted hadronic $W$ tagging}
\label{sec:wtag}

As mentioned in Sec.~\ref{sec:intro}, highly lorentz boosted, hadronically decaying $W$ bosons commonly arise in extensions of the Standard Model.  The boost causes the decay products of these bosons to be collimated in the lab frame and to be  mostly captured by a single large-radius jet.  Various features of the substructure of these jets can be used to distinguish the boosted bosons from generic quark and gluon jets.  

A bump hunt is performed either in the mass of the $W$ candidate jet, $m_J$, or the mass of one $W$ candidate jet and another (possibly $W$ candidate) jet, $m_{JJ}$.  The challenge with substructure classifiers is that they can introduce artificial bumps into the mass spectrum because substructure is correlated with the jet mass and the jet kinematic properties (which are related to $m_{JJ}$).  For this reason, boosted $W$ tagging has become a benchmark process for studying decorrelation methods at the LHC.

The simulated samples used in this section are the same as in Ref.~\cite{DiscoFever} (intended to emulate the study in Ref.~\cite{ATL-PHYS-PUB-2018-014}) and are briefly summarized here.  In particular, boosted $W$ bosons (signal) and generic multijet (background) events are generated with \textsc{Pythia} 8.219~\cite{Sjostrand:2006za,Sjostrand:2014zea} and a detector simulation is provided by \textsc{Delphes} 3.4.1~\cite{deFavereau:2013fsa,Mertens:2015kba,Selvaggi:2014mya}.  Jets are clustered using the anti-$k_t$ algorithm~\cite{Cacciari:2008gp} with $R=1.0$ implemented in \textsc{FastJet} 3.0.1~\cite{Cacciari:2011ma,Cacciari:2005hq}.  The selected jets for this study have 300~GeV~$<p_T<$~400~GeV and 50~GeV$<m_J<$~300~GeV.  Ten representative jet substructure features are computed for each jet and used for classification.  This list is the same as in Ref.~\cite{ATL-PHYS-PUB-2018-014} (based on Ref.~\cite{ATLAS-CONF-2017-064}) and includes the energy correlation ratios $C_2$ and $D_2$~\cite{Larkoski:2014gra}, the $N$-subjettiness ratio $\tau_{21}$~\cite{Thaler:2010tr}, the Fox-Wolfram moment $R_2^\text{FW}$~\cite{PhysRevLett.41.1581}, planar flow $\mathcal{P}$~\cite{Almeida:2008tp}, the angularity $a_3$~\cite{Aad:2012meb}, aplanarity $A$~\cite{Chen:2011ah}, the splitting scales $Z_\text{cut}$~\cite{Thaler:2008ju} and $\sqrt{d_{12}}$~\cite{Aad:2013ueu}, and the $k_t$ subjet opening angle $KtDR$~\cite{CATANI1993187}. Detailed explanations of these features can be found in the references.

\subsubsection{Classifier Details}

\paragraph{MoDe and DisCo:}
We use a simple 3-layer neural network with a similar architecture to that described in Ref.~\cite{DiscoFever}. 
However, unlike Refs.~\cite{DiscoFever} and \cite{ATL-PHYS-PUB-2018-014}, after each of the 3 fully connected 64-node layers, we use Swish activation~\cite{ramachandran2017searching} as it provides a notable performance increase. 
We also use a batch normalization layer after the first fully connected layer. 
The output layer has a single node with a sigmoid activation. 
Both \textsc{MoDe} and DisCo are trained with the ADAM optimizer~\cite{DBLP:journals/corr/KingmaB14} using a 1cycle learning rate policy~\cite{smith2018superconvergence} with a starting learning rate of $10^{-3}$ and a maximum learning rate ($lr$) of $10^{-2}$, which is reached using a cosine annealing strategy~\cite{loshchilov2017sgdr} and decayed to $10^{-5}$ during the last few iterations. 
Momentum is cycled in the inverse direction from 0.95 to a minimum of 0.85.
These hyperparameters were selected through a learning rate range test.
Training is done using large batches of 10--20\% of the training data. 
Note that large batch sizes do not necessarily make training more difficult especially when coupled with the 1cycle learning policy; see Ref.~\cite{smith2017superconvergence}.

\paragraph{Adversarial Decorrelation:}
The same classifier used for {\sc MoDe} and DisCo is trained against a Gaussian Mixture Network (GMN)~\cite{Bishop94mixturedensity} that parametrizes a Gaussian mixture model with 20 components, {\em i.e.}\ its outputs are the means, variances, and mixing coefficients of 20 normal distributions.
We follow a similar adversarial setup to that referenced in Refs.~\cite{ATL-PHYS-PUB-2018-014} and \cite{DiscoFever}. 
We use one hidden layer with 64 nodes with ReLu activation connected to 60 output nodes. 
These outputs are then used to model the posterior probability density function  $p_{\theta_{\mathrm{adv}}}(m|f(X)=s)$, where $\theta_\mathrm{adv}$ are the parameters of the GMN. 
The adversary is optimized by maximizing the likelihood of the data given by
\begin{equation}
    L_\mathrm{adv} = \mathbb{E}_{s\sim f(X)}\mathbb{E}_{m\sim M|s} \left[- \log\, p_{\theta_\text{adv}}(m|s)\right].
\end{equation}
The training procedure starts by training the classifier alone for 20 epochs with $lr=10^{-4}$ followed by 20 epochs of adversarial training only with $lr=5\cdot10^{-3}$. 
Finally, both networks are trained simultaneously by optimizing the following loss function 
\begin{align}
    \arg \min_{\theta_{\mathrm{class}}}\;  \max_{\theta_{\mathrm{adv}}} \left[L_{\mathrm{class}(\theta_\mathrm{class})} - \lambda L_{\mathrm{adv}}(\theta_\mathrm{class}, \theta_\mathrm{adv})\right],
    \end{align}
where $\theta_\mathrm{class}$ and $\theta_\mathrm{adv}$ are the parameters of the classifier and the adversary, respectively. 
To control the classification-decorrelation trade-off, we vary $\lambda$ between 1 and 100. 
The non-convex nature of the loss makes training considerably more difficult; the hyperparameters must be chosen carefully.


\subsubsection{Decorrelation}

First, we will show that {\sc MoDe}[0] achieves state-of-the-art decorrelation performance.
Following Ref.~\cite{DiscoFever}, we quantify the classification and decorrelation performance using the following metrics:
R50, the background rejection power (inverse false positive rate) at $50\%$ signal efficiency;
and 1/JSD, where the Jensen-Shannon divergence (JSD)
is a symmetrized version of the Kullback–Leibler divergence. 
Here, JSD is used to compare the mass distributions of backgrounds that pass and fail the classifier-based selections, 
with the relative entropy measured in bits.

\begin{figure}[t]
\centering
\includegraphics[width=0.49\textwidth]{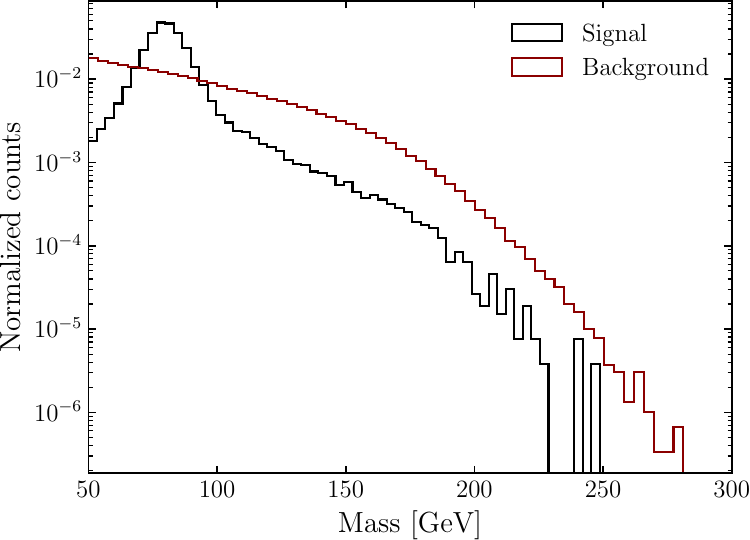}
\includegraphics[width=0.49\textwidth]{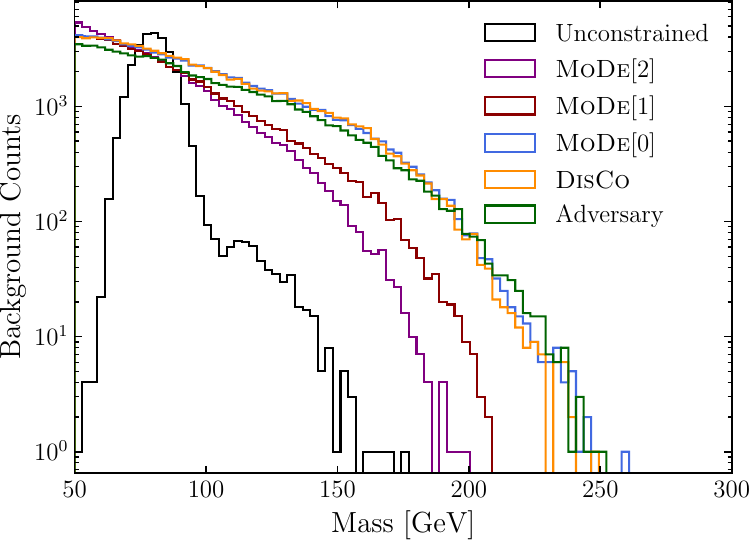}
\caption{{\bf Left:}
    Distributions of signal and background events without selection. 
  {\bf Right:}
  Background distributions at 50\% signal efficiency (true positive rate) for different classifiers. The unconstrained classifier sculpts a peak at the $W$-boson mass, while other classifiers do not.}
  \label{fig:wtagdata}
\end{figure}

\begin{figure}
    \centering
    \includegraphics[width=0.7\linewidth]{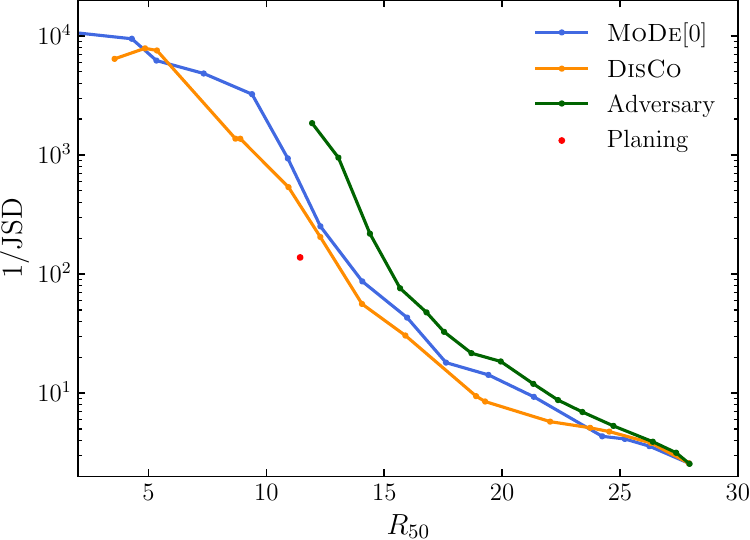}
    \caption{Decorrelation versus background-rejection power showing that {\sc MoDe}[0] performs similarly to existing state-of-the-art decorrelation methods.}
    \label{fig:phase}
\end{figure}

Figure~\ref{fig:wtagdata} shows that, as expected, without imposing a strong constraint on mass decorrelation, the classifier learns to select samples near the $W$-boson mass, which sculpts a fake peak in the background.
Figure~\ref{fig:wtagdata} also shows that {\sc MoDe}[0] successfully decorrelates its response from mass (if the decorrelation hyperparameter $\lambda$ is chosen to be sufficiently large). 
Figure~\ref{fig:phase} shows that the existing state-of-the-art decorrelation methods discussed in Sec.~\ref{sec:methods} perform similarly to {\sc MoDe}[0] on this $W$-tagging problem.
More precisely, as observed in Ref.~\cite{DiscoFever}, the adversary method performs slightly better, but is considerably more difficult to train, since 
it requires carefully tuning two neural networks against each other. 
The optimal solution is a saddle point, where the classification and adversarial losses are minimized and maximized, respectively, which makes the training inherently unstable.
Conversely, both DisCo and {\sc MoDe}[0] minimize convex loss functions, making their training robust and stable, and both only introduce one additional hyperparameter in the loss function.
The performances of {\sc MoDe}[0] and DisCo are comparable in these metrics (Appendix~\ref{app:optim} shows that {\sc MoDe}[0] is less resource intensive), though decorrelation is not our primary objective. 

\subsubsection{Beyond Decorrelation}

Moving beyond decorrelation the 1/JSD metric is no longer relevant.
Figure~\ref{fig:wtagdata} shows that neither {\sc MoDe}[1] nor {\sc MoDe}[2] sculpts a peaking structure in the background, but their 1/JSD values are small since neither seeks to decorrelate from the mass. 
Therefore, we replace the 1/JSD metric with the signal bias induced by the classifier selection, which is what actually matters when searching for resonant new physics.
Specifically, we use the signal estimators obtained by fitting the selected background-only samples to a simple polynomial function as proxies for the signal biases.
These are divided by their uncertainties such that values of roughly unity are consistent with no bias, while values significantly larger than unity indicate substantial bias that could result in false claims of observations.

\begin{figure}[t]
\centering 
\includegraphics[width=0.7\textwidth]{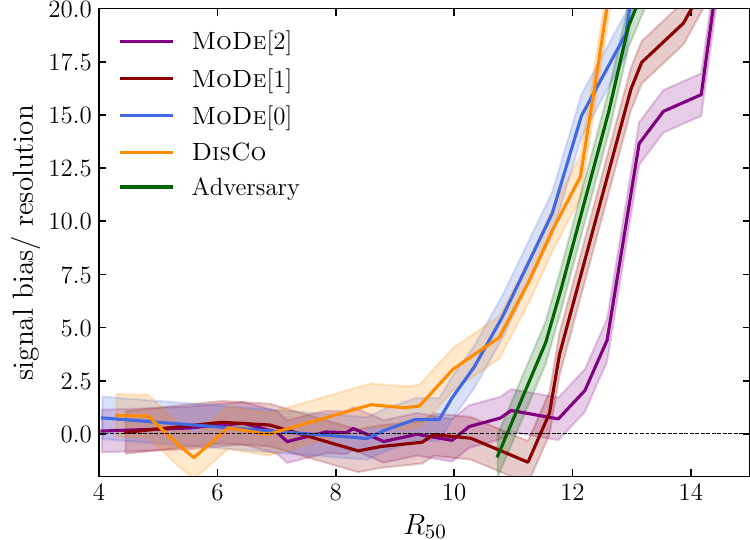}
\caption{Signal bias relative to resolution, which is roughly the square root of the background in the signal region, versus background-rejection power. The flexibility beyond simple decorrelation provided by {\sc MoDe}[1] and {\sc MoDe}[2] result in improved performance, {\em i.e.}\ larger rejection power.}
\label{fig:bias}
\end{figure}

Figure~\ref{fig:bias} shows that the DisCo and {\sc MoDe}[0] decorrelation methods provide unbiased signal estimators for ${\rm R50} \lesssim 9$,  which from Fig.~\ref{fig:phase} corresponds to $1/{\rm JSD} \gtrsim 1000$. 
While achieving higher decorrelation values is possible, this does not provide any tangible gains in the bump-hunt analysis.
Figure~\ref{fig:bias} also shows that the flexibility to go beyond decorrelation provided by {\sc MoDe}[1] and {\sc MoDe}[2] results in achieving unbiased signal estimators at larger background-rejection power.
This would directly translate to improved sensitivity in a real-world analysis. 
For example, since it is likely that only unbiased classifiers would be considered, Fig.~\ref{fig:bias} can be used to estimate the improvement in the signal cross-section sensitivity for the $W$-tagging analysis, which scales roughly like $\sqrt{R_{50}}$, using the classifier of each type with the largest $R_{50}$ value that is consistent with being unbiased. 
{\sc MoDe}[1] and {\sc MoDe}[2] provide roughly 5\% improved sensitivity over the adversary method, which recall is considerably more difficult to train, and 10--20\% improvements over the other decorrelation methods. 
We note, however, that how much is gained will strongly depend on the specifics of the problem, {\em e.g.}, how large of a mass range is considered and whether the signal mass is known or if a scan in mass will be done.

\section{Conclusions and Outlook}
\label{sec:conclusion}

In summary, a key challenge in searches for resonant new physics is that classifiers trained to enhance potential signals must not induce localized structures.  
In particular, if classifiers can infer the mass of the parent resonance, then selecting signal-like events will simply pick out background events with a reconstructed mass near the target resonance mass creating an artificial structure in the background.
Such structures could result in a false signal when the background is estimated from data using sideband methods.  
A variety of techniques have been developed to construct classifiers which are independent from the resonant feature (often a mass).  
Such strategies are sufficient to avoid localized structures, but are not necessary.

In this article, we presented a new set of tools using a novel moment loss function (Moment Decomposition or \textsc{MoDe}) which relax the assumption of independence without creating structures in the background.  
Using \textsc{MoDe}, analysts can require independence, linear dependence,  quadratic dependence, {\em etc}. 
In addition, analysts can place bounds on the slope of the linear dependence, and restrict higher-order dependence to be monotonic.
By allowing classifiers to be more flexible, we enhance the sensitivity to new physics without compromising the fidelity of the background estimation.

\section*{Code and Data}

An implementation for \textsc{MoDe} in PyTorch as well as Keras/Tensorflow is available at \url{https://github.com/okitouni/MoDe}, along with example code used to produce the results in this article.  The simulated $W$ and QCD samples are available from Zenodo at Ref.~\cite{gregor_kasieczka_2020_3606767}.

\section*{\label{sec::acknowledgments}Acknowledgments}

We thank David Shih and Gregor Kasieczka for providing the simulated $W$ and QCD examples from Ref.~\cite{DiscoFever} and for making their code public.  We also thank David Shih and Jesse Thaler for helpful conversations and feedback on the manuscript.
OK was supported by an MIT fellowship.
BPN was supported by the Department of Energy, Office of Science under contract number DE-AC02-05CH11231.
CW and MW were supported by the National Science Foundation under contracts PHY-1912836 and OAC-1739772. In addition, this work was supported by the National Science Foundation under Cooperative Agreement PHY-2019786 (The NSF AI Institute for Artificial Intelligence and Fundamental Interactions, http://iaifi.org/).

\clearpage 

\begin{appendices}
\section{Scalability}
\label{app:optim}
\begin{figure}[ht]
\centering
\begin{minipage}[b]{0.49\linewidth}
\includegraphics[width=1.1\textwidth]{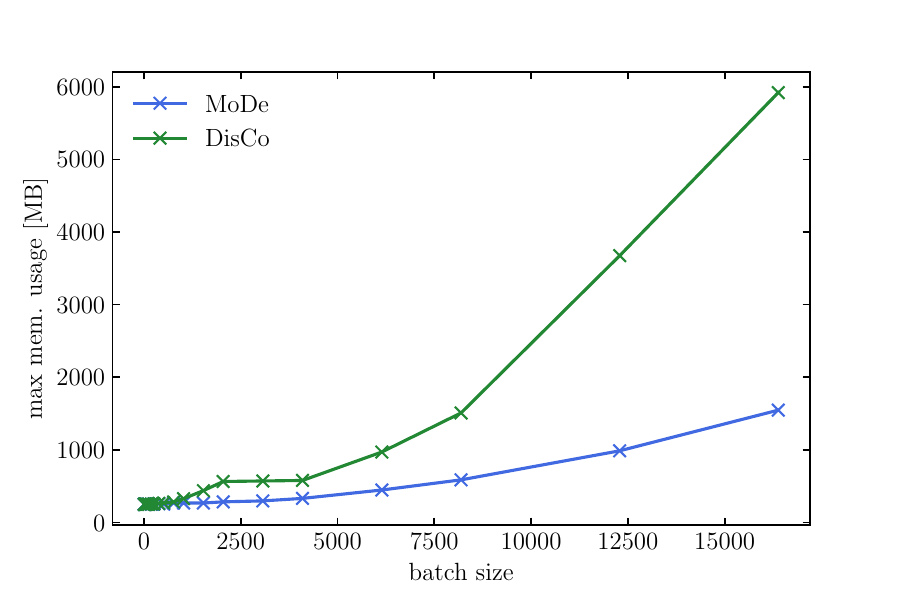}
    \label{fig:minipage1}
\end{minipage}
\begin{minipage}[b]{0.49\linewidth}
    \includegraphics[width=1.1\textwidth]{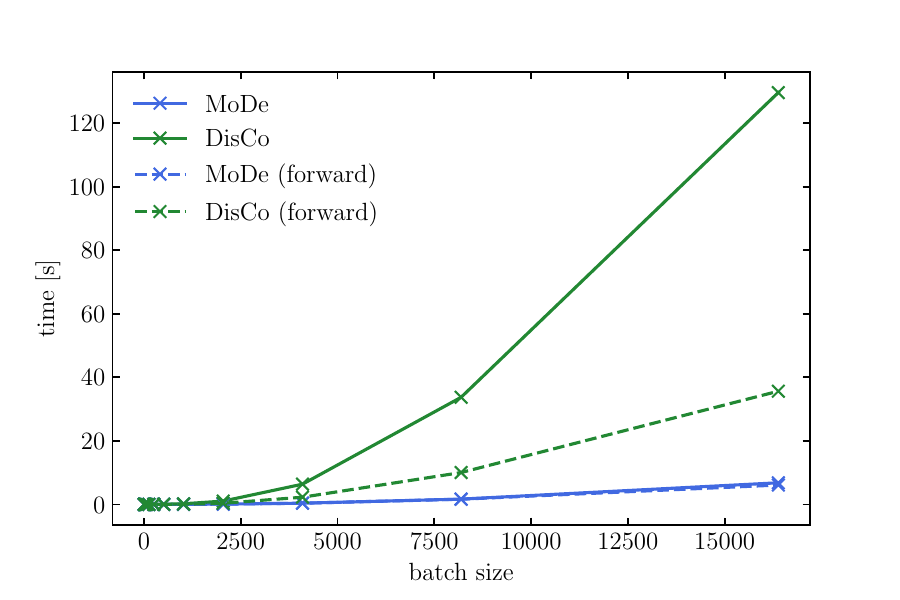}
    \label{fig:minipage2}
    \end{minipage}
\vspace*{-20pt}
\caption{Maximum memory usage and CPU time for different batch sizes of triples $(s_i,m_i,y_i)$.}
\end{figure}

\end{appendices}

\bibliographystyle{JHEP}
\bibliography{main,HEPML}

\end{document}